\documentclass[12pt,letterpaper]{article}
\usepackage[utf8]{inputenc}
\usepackage[margin=1in]{geometry}   
\usepackage{fourier}                
\usepackage{natbib}
\usepackage{xcolor}
\usepackage[final]{pdfpages}
\usepackage[colorlinks]{hyperref}
\usepackage[affil-it]{authblk}
\hypersetup{
citecolor = {blue}
}

\usepackage{graphicx}
\begin{document}

\includepdf[pages=-]{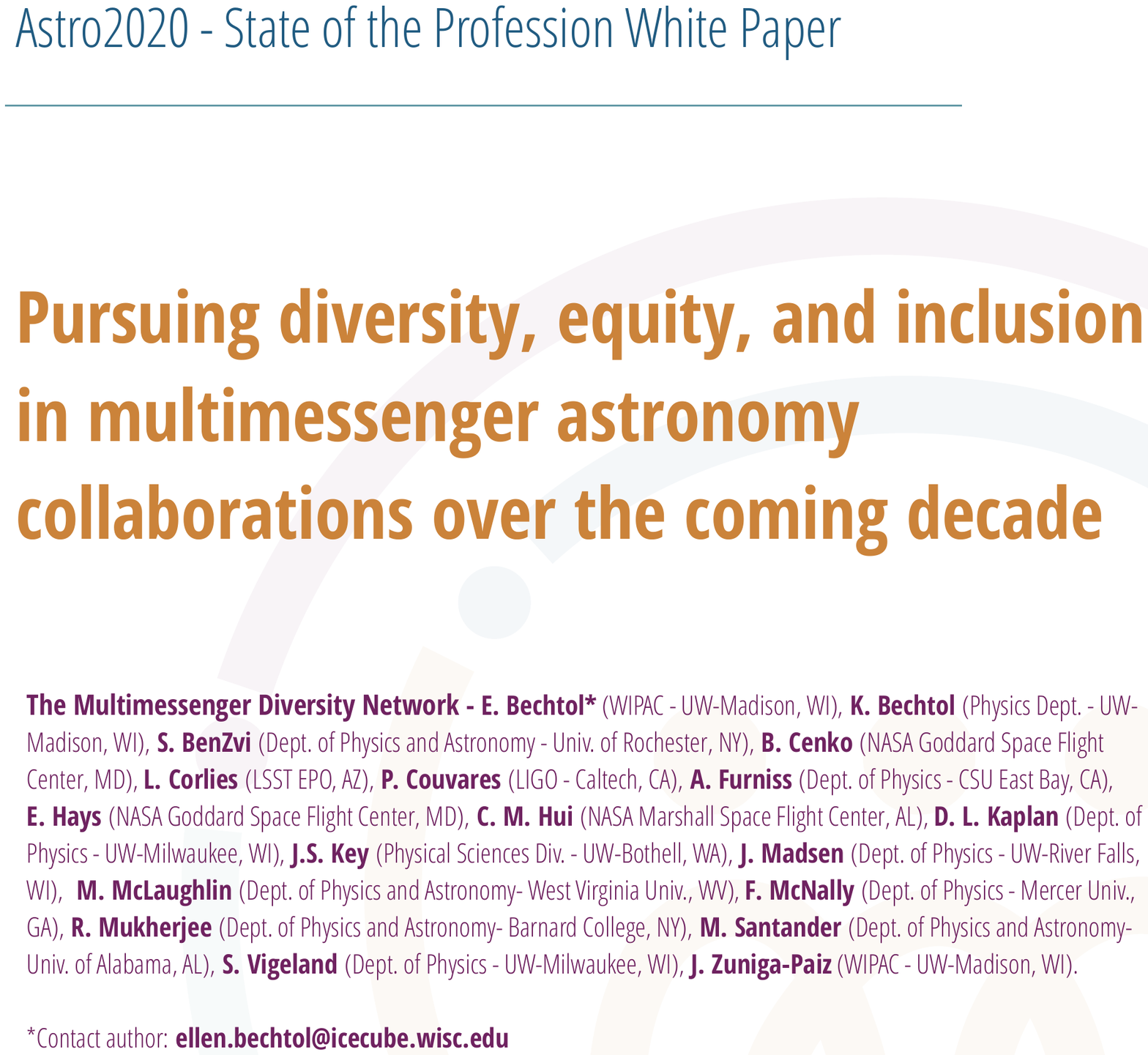} 

\clearpage
\setcounter{page}{1}

\newpage


\noindent{\textbf{\large Abstract}}

A major goal for the astronomy and astrophysics communities is the pursuit of diversity, equity, and inclusion (DEI) in all ranks, from students through professional scientific researchers.
Large scientific collaborations \textemdash{} increasingly a primary place for both professional interactions and research opportunities \textemdash{} can play an important role in the DEI effort. 
Multimessenger astronomy, a new and growing field, is based on the principle that working collaboratively produces synergies, enabling advances that would not be possible without cooperation. The nascent Multimessenger Diversity Network (MDN) is extending this collaborative approach to include DEI initiatives. After we review of the current state of DEI in astronomy and astrophysics, we describe the strategies the MDN is developing and disseminating to support and increase DEI in the fields over the coming decade:
\par
\begin{itemize}
    \item Provide opportunities (real and virtual) to share DEI knowledge and resources
    \item Include DEI in collaboration-level activities, including external reviews
    \item Develop and implement ways to recognize the DEI work of collaboration members
\end{itemize}
    

\section{Introduction} 

There are multiple motivations for addressing the well-documented lack of diversity in astronomy, physics, and astrophysics evident in educational institutions and scientific research collaborations \citep{Merner:2017aip, Porter:2019aip,NCSES18}. Broadening participation benefits the field at large, since more diverse collaboration improves scientific output \citep{Chubin:2008,Page:2007}. Moreover, increased diversity is important from an ethical and social justice perspective.   

There are also significant issues that slow DEI progress. Most diversity, equity, and inclusion (DEI) work in research collaborations is volunteer-based. The lack of time, professional regard, and explicit job responsibility for DEI are often anecdotally cited by collaboration members as reasons for not getting involved with DEI work. This is similar to experiences with education and outreach (E\&O) where reasons for not contributing have been probed. For example, in a survey of 131 members of the Dark Energy Survey, respondents indicated they would be encouraged to get involved with E\&O work if they felt supported to allocate time to it (53\%), if E\&O were an explicit part of their job duties (46\%), and if E\&O were more highly regarded among peers (39\%) \citep{Farahi:2019}.  The same factors likely hinder the sustained participation in DEI efforts. In addition, people from underrepresented groups face hurdles related to implicit bias in hiring \citep{Eaton2019} and retention.

Science efforts are coalescing to bring astrophysicists and astronomers together to better explore and understand the cosmos, leveraging the work and expertise of many different experiments and facilities. Multimessenger astronomy, one of the \href{https://www.nsf.gov/news/special_reports/big_ideas/}{10 Big Ideas} for future funding from the National Science Foundation (NSF), points to the promise of multiple views of the same phenomena in producing results not possible without cooperative, complementary research. Multimessenger astronomy provides a real reason for interactions and rewards those who participate with greater science productivity. We believe a similar approach can be applied to address DEI issues in astronomy and astrophysics, leveraging  DEI experiences and involvement to create more lasting and meaningful change than what is possible through individual actions. 

\begin{figure}
\begin{center}
  \includegraphics[width=4in]{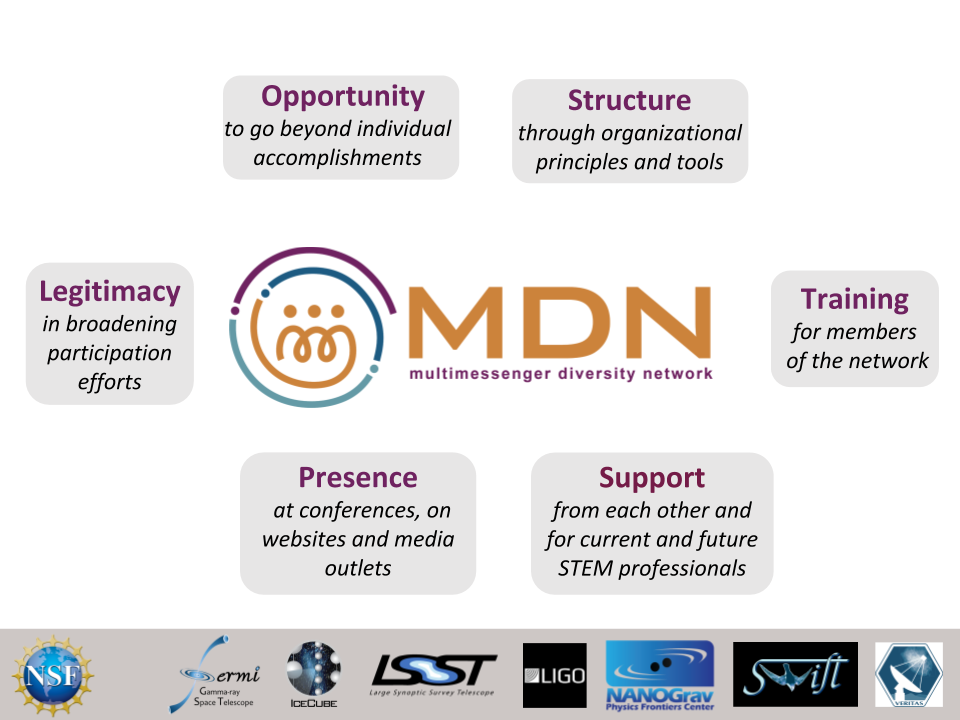}
  \caption{The elements of the MDN and current members, the Fermi Gamma-ray Space Telescope, IceCube Collaboration, Large Synoptic Survey Telescope (LSST), Laser Interferometer Gravitational-Wave Observatory (LIGO), Nanohertz Observatory for Gravitational Waves (NANOGrav), Neil Gehrels \emph{Swift} Observatory, and Very Energetic Radiation Imaging Telescope Array System (VERITAS).}
  \label{fig:MDN}
  \end{center}
\end{figure}

One such effort is the Multimessenger Diversity Network (MDN), an NSF INCLUDES-funded community of representatives from multimessenger research collaborations, focused on increasing diversity by sharing knowledge, experiences, training, and resources. One of the NSF's 10 Big Ideas, INCLUDES (Inclusion across the Nation of Communities of Learners of Underrepresented Discoverers in Engineering and Science) focuses on transforming education and career pathways to help broaden participation in science and engineering. The goal of the MDN is to broaden participation in multimessenger astronomy, including identifying current and past successful initiatives led or joined by the multimessenger community. Members of the MDN will also become the DEI engagement fellows at their experiment, with the goal of implementing best practices and new strategies for action. The elements and current members of the MDN are shown in Figure~\ref{fig:MDN}.

The MDN will leverage existing communication channels to expand the very successful research collaboration of multimessenger observatories. 
The initial recommendations of the MDN, based on our experiences as a community and our  perspectives as members of many different multimessenger research collaborations, are three strategies to support sustainable DEI efforts: 
\begin{itemize}
    \item Provide opportunities to share knowledge and resources on DEI-related activities
    \item Include DEI in collaboration-level activities, including external reviews
    \item Develop and implement ways to recognize the DEI work of collaboration members
\end{itemize}
\par


\section{Diversity and Current State of Affairs}

\begin{figure}
\begin{center}
  \includegraphics[width=4in]{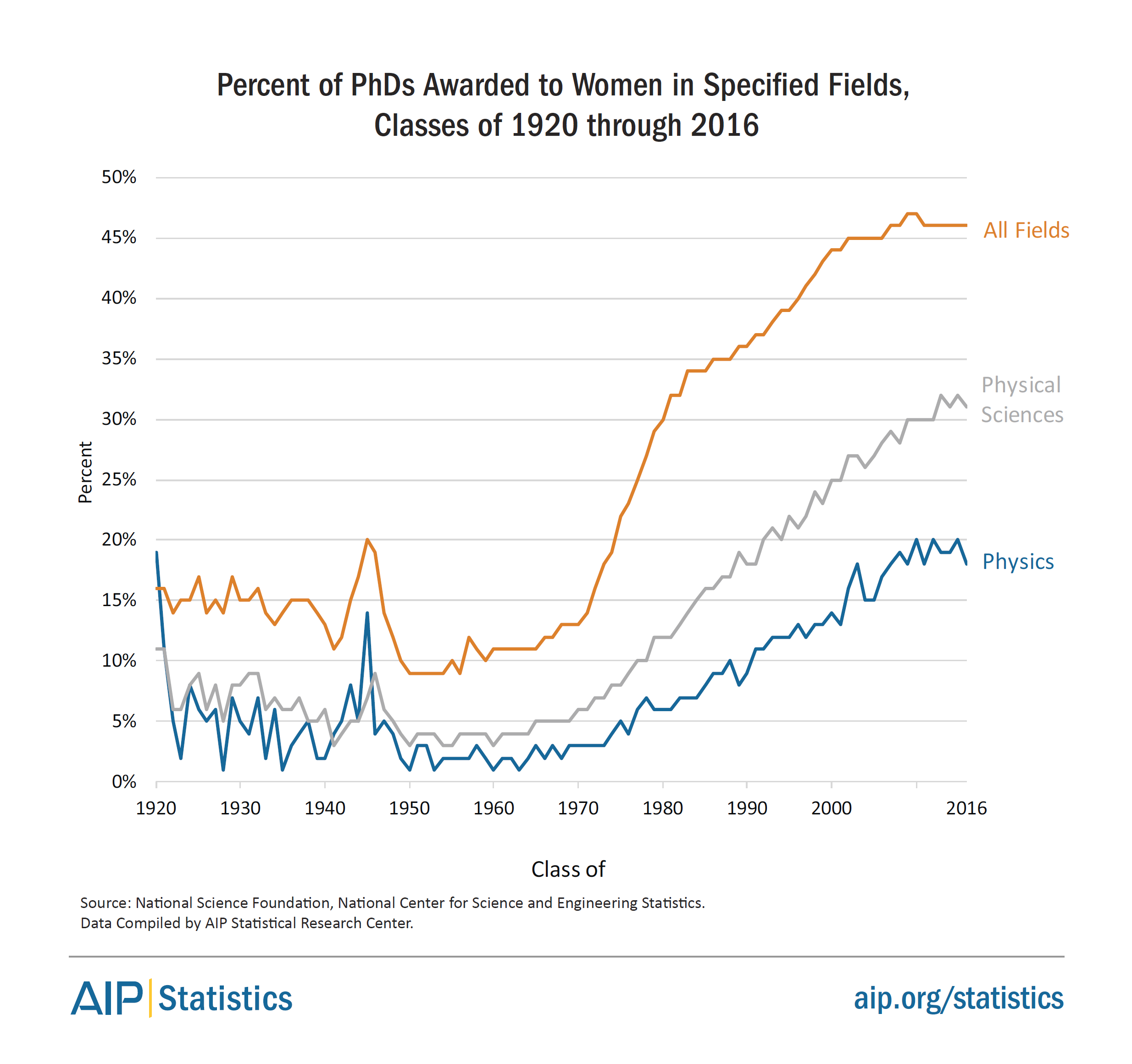}
  \caption{Percentage of PhDs awarded to women in physics and astronomy, compared to other sciences (from \cite{Porter:2019aip}). }
  \label{fig:womenstem}
  \end{center}
\end{figure}

The American Institute of Physics (AIP) has gathered demographic data on the astronomy and physics communities for many decades through their \href{https://www.aip.org/statistics}{Statistical Research Center}. For example, Fig. \ref{fig:womenstem} shows the percentage of PhDs earned by women over roughly the last century.  From 1920 to 1950, women in physics and astronomy earned roughly the same percentage of PhDs as their counterparts in the other physical sciences, at about one half to one third the rate of all PhDs granted to women in this period. Since 1950, there has been growth in the overall percentage of female PhDs across all disciplines, but the gains in astronomy and physics are well behind the rest of the physical sciences. This disparity is seen at other levels as well: while there has been an increase in the participation of women in physics since the 1920s, the number of women at the postdoctoral or faculty level is still below that of other disciplines \citep{Porter:2019aip}. According to the National Center for Education Statistics (2016), women earned over 50\% of all bachelor’s degrees, but only 21\% of physics and 33\% of astronomy bachelors’ degrees. \citep{Porter:2019aip}. 

Other reports have documented small percentages and slow growth in representation of minorities in the sciences. The Survey of Earned Doctorates is an annual census of individuals who attain doctorates from accredited U.S academic institutions. This survey is sponsored by six federal agencies, including the NSF. The most comprehensive publication in this report is the \textit{Doctorate Recipients from U.S. Universities}, which measures the steady investment in STEM and scholarship as well as the increased representation of women and underrepresented minorities in various fields. The 2018 report demonstrates that in the fields of physics and astronomy, Hispanic/Latinx, Black/African American, and Indian/Alaskan Native scholars make up only 5.3\%, 2\%, and 0\%, respectively, of physics and astronomy doctoral degrees received in 2017 compared to their overall percentages in the US population of 18\%, 13\%, and 1.3\% \citep{NSF18}. These percentages demonstrate that despite incremental growth throughout the years, there is still a deficit of diversity in the astronomy and physics communities. 


\section{Roles for Large Scientific Collaborations}


%

An increasing fraction of astronomers participate in large international scientific collaborations and, in many cases, have daily interactions with colleagues from institutions around the world.
For many astronomers, affiliation with a scientific collaboration has become a significant part of their professional life and identity, which persists across changes in institutional affiliation, supervisor, and career stage.
Scientific collaborations can directly contribute to DEI efforts by providing access to new and comprehensive datasets and research opportunities, supplying mentors at multiple career stages, and establishing policies that recognize and promote the contributions of all members (e.g., organization structure, speaker's bureau, publication boards).
Most importantly, scientific collaborations can accelerate the exchange of best practices between institutions or departments.
At the same time, large scientific collaborations are limited in the degree that they can directly affect change at individual institutions, e.g., through hiring and tenure practices.

\section{Recommendations}

\subsection{Providing opportunities for shared knowledge}
Astronomers have individually recognized the need for increased participation in diversity and inclusion efforts, particularly within the last decade \citep{Norman:2009,Norman:2009b,Nota:2009,Waller2009,Rudolph19}. There are astronomy programs with DEI initiatives led by full-time faculty, staff, and groups of graduate students (e.g., \href{https://sites.google.com/umich.edu/astro-dei/}{Astro DEI}). Professional organizations and affiliated groups have also devoted resources and volunteer time towards DEI issues (e.g., \href{https://lgbtphysicists.org/}{LGBTphysicists.org}, \href{https://cswa.aas.org/}{AAS Committee on the Status of Women}, \href{https://www.aps.org/about/governance/committees/commin/index.cfm}{Committee on Minorities in Physics}). Similarly, many scientific research collaborations are undertaking DEI efforts but with few ways to share their experiences and support each other; the MDN is a rare counterexample. While localized efforts are powerful on small scales, there are natural opportunities to share experiences, leverage developed materials, and pool resources both online and in person. As a network of large multimessenger projects, we recommend the following actions be taken by our respective groups and considered as a model to be adopted more generally:  



\begin{itemize}
 \setlength{\itemsep}{0em}

 \item Support the creation of local or virtual DEI groups for more consistent, organized efforts throughout the year. 
 
 \item Support opportunities to collaborate and co-develop new DEI efforts.
 
 \item Support mentoring opportunities within the multimessenger community, either within individual collaborations or across collaborations.
 
 \item Develop an open, shared database of ongoing DEI efforts.
  \end{itemize}
  
An open database would allow for the easy sharing of materials such as codes of conduct, antiharassment policies, and climate surveys. It would also create an extended network among those interested in and working to create inclusive and welcoming collaborations. In this way, existing programs can easily be extended and new ideas successfully launched, all while learning from and supporting each other. 

As a way of jumpstarting this effort, we have created a mailing list
\footnote{\href{https://groups.google.com/forum/\#!forum/multimessenger-diversity-network}{https://groups.google.com/forum/\#!forum/multimessenger-diversity-network}} 
for the MDN that other collaborations, groups, or individual members interested in DEI efforts can freely join to exchange information and experiences. Materials collected and generated by the MDN as well as progress on a future database will be circulated through the mailing list.

\subsection{Include diversity, equity, and inclusion in collaboration-level activities.}

Collaboration-level activities such as onboarding for new members, recurring virtual meetings, and regular face-to-face meetings are opportunities to create inclusive and welcoming environments. We recommend that collaborations include these best practices:

 \begin{itemize}
 \item Host DEI sessions at project-led community workshops and conferences. 
 
 \item Promote DEI work in external reviews through discussions of broader impacts.
 
 \item Give DEI contributions equal status to science talks, including a DEI-focused plenary talk.
 
 \item  Make meetings inclusive. 
 \end{itemize}
 
 There are several resources available on how to host inclusive and accessible meetings (e.g., \href{https://discover-cookbook.github.io/}{DISCOVER cookbook} and \href{https://sparcopen.github.io/opencon-dei-report/}{OpenCon}). Some examples of actions that can be taken include adopting and enforcing a code of conduct, providing childcare assistance, moderating panel Q\&A sessions to give voice to early career scientists and people from underrepresented groups, and considering the accessibility needs of attendees.

\subsubsection{Case Study: NANOGrav}

All NANOGrav collaboration meetings have a session devoted to training on a diversity-related topic, presented by an external expert, that all attendees are expected to attend. NANOGrav collaboration members who are not present must watch a recording of that session in order to renew their membership. In addition, meeting session chairs are trained in best practices to involve students, including waiting to take questions by senior personnel until after student questions have been asked. The collaboration also prepares and engages students via  pre-meeting workshops and multiple debriefing sessions.

\subsection{Recognition of diversity and inclusion work}
Individuals who support DEI efforts for their projects and their universities are often unrecognized and undervalued, even for positive outcomes requiring a large time investment. Research collaborations should develop and share ways to recognize DEI work performed by members, for example, by taking the following steps:  
\begin{itemize}
\item Include DEI in collaboration-level reviews for membership, authorship, and service. 

\item Celebrate individual or group DEI efforts within the collaboration through named prizes and/or awards.

\item Educate departments, funding agencies, and hiring committees about the value of DEI and encourage them to consider DEI efforts in decision making. 
\end{itemize}

Some of this work is happening already \textemdash{} the LIGO, NANOGrav, and LISA collaborations currently recognize E\&O and DEI work as collaboration-level service \textemdash{} but further steps can be taken. Examples include providing press releases and social media recognition for DEI efforts and including formal letters outlining DEI contributions for tenure reviews.  



\section{Assessment} 
Assessment and evaluation of DEI efforts is an important component in broadening participation and creating welcoming environments. To start, collaborations need to understand their current culture from the perspectives of members at all career levels and backgrounds. A survey or interviews with a diverse set of collaboration members from different backgrounds and career stages can give a good sense of how inclusive the culture is. This work should be done with some regularity and  with the results distributed to collaboration members and leadership. Research collaborations can track how they implement the recommendations in this white paper and subsequent impacts through note-taking and self-reflection. For example, one method would be implementing a debriefing activity after each collaboration meeting, with a focus on inclusive practices, or writing a yearly report on DEI efforts in order to provide regular benchmarks and points for reflection. Assessment and evaluation is another area where collaboration within the field has promising benefits, as in-house expertise and experience in DEI assessment is often lacking. Following best practices from other fields, looking to large well-funded efforts, and sharing experiences and resources are likely ways forward. 

\section{Summary}
Making the astronomy and astrophysics communities more diverse, equitable, and inclusive is a big challenge. Immediate steps can be taken to address environmental issues that contribute to the homogeneity of the fields. Attracting and retaining top talent is essential for viability of any discipline.  If large portions of the population are not participating, it is likely that opportunities for advances are being missed.  The MDN is bringing together people from multimessenger astronomy to collectively address DEI issues.  This short note reviewed the current state and provided some specific actions that can be taken to improve DEI in the astronomy and astrophysics communities. The goal is to build sustainable DEI programs that ensure a better working environment for all.        

\bibliographystyle{apalike}

\end{document}